\documentclass[aps,pre,twocolumn,groupedaddress,showpacs]{revtex4}
\usepackage{graphicx}
\newcommand{\bv}{\mathbf{v}}

\newcommand{\bR}{\mathbf{R}}
\newcommand{\bbr}{\mathbf{r}}

\newcommand{\sep}{ \ \ \ , \ \ \ }

\newcommand{\beq}{\begin{equation}}
\newcommand{\eeq}{\end{equation}}
\newcommand{\beqn}{\begin{eqnarray}}
\newcommand{\eeqn}{\end{eqnarray}}
\newcommand{\pp}{\partial}
\newcommand{\dd}{{\rm d}}
\newcommand{\ee}{{\rm e}}
\newcommand{\ii}{{\rm i}}
\newcommand{\eq}{Eq.\ }
\newcommand{\eqs}{Eqs\ }
\newcommand{\fig}{Fig.\ }

\newcommand{\la}{\langle}
\newcommand{\ra}{\rangle}

\newcommand{\cO}{{\cal O}}

\newcommand{\cA}{U}

\newcommand{\cf}{c.f.\ }

\newcommand{\hrho}{\hat{\rho}}

\newcommand{\sect}{{\rm Sect.\ }}
\newcommand{\erf}{{\rm erf}}

\begin{document}

\title{Fluctuation-induced collective motion:
 A single-particle density analysis
}
\author{Chiu Fan Lee
}
\email{cflee@pks.mpg.de}
\affiliation{Max Planck Institute for the Physics of Complex Systems,
N\"{o}thnitzer Str.~38, 01187 Dresden,
Germany
}


\date{\today}

\begin{abstract}
In a system of noisy self-propelled particles with interactions that favor directional alignment, collective motion will appear if the density of particles increases 
beyond a certain threshold. In this paper, we argue that such a threshold may depend also on the profiles of the perturbation in the particle directions. Specifically, we 
perform mean-field, linear stability, perturbative and numerical analyses on an approximated form of the Fokker-Planck equation describing the system. We find that if 
an angular perturbation to an initially homogeneous system  is large in magnitude and  highly localized in space, it will be amplified and thus serves as an indication of the onset of collective motion. Our results also demonstrate that high particle speed promotes collective motion.
\end{abstract}

\maketitle
\section{Introduction}
The interesting phenomena of flocking in animals 
\cite{Couzin_Nature05, Buhl_Science06, Sumpter_PRSB06} and self-organized patterns in motile cells \cite{Tsimring_PRL95,Riedel_Science05,Budrene_Nature91} are currently 
driving the 
intense theoretical study of collective motion among self-propelled particles 
\cite{Vicsek_PRL95,Toner_PRL95, Toner_PRE98,Gregoire_PRL04,Dossetti_PRE09,Romanczuk_PRL09,Aldana_PRL07,DOrsogna_PRL06,Kruse_PRL04, 
Bertin_PRE06,Peruani_EPJST08}. In particular, a comprehensive linear stability analysis on the onset of collective motion from the perspective of Boltzmann equation has 
recently appeared \cite{Bertin_JPA09}. 
 Models for collective motion usually involve motile particles that possess alignment interactions and angular noise. Collective motion is then observed  if the density 
of particles increases beyond a 
certain threshold.
Besides density fluctuations, fluctuations in the heading directions of the particles constitute another important aspect of the system. 
Here, we study a minimal model for collective motion and  show that the threshold for collective motion transition may depend on the profiles of directional 
fluctuations.
 Specifically, we find that an initial directional perturbation to a spatially homogeneous system  will be amplified, if  
the perturbation is large in magnitude and is highly localized in space. We also demonstrate that high particle speed promotes collective motion.

To achieve our 
results, we first write down the Fokker-Planck equation describing the single-particle density distribution of the system in \sect \ref{sModel}. We then investigate in \sect \ref{sMF} the 
equation in the Fourier space 
of the directional component, and argue that only the lower order modes are important at the onset of collective motion. As a result, the 
dynamics of the distribution function can be captured by a set of three nonlinear coupled differential equations, which we subsequently study with linear stability analysis in 
\sect \ref{sLSA}. In \sect \ref{sLast}, we go beyond the linear stability regime by investigating the dynamical equations perturbatively and numerically.

\section{Model}
\label{sModel}
In this work, we follow \cite{Peruani_EPJST08} and consider a minimal model for collective motion in two dimensions, where every particle is assumed to have constant 
speed and that their interactions consist only of directional alignment mechanism. Noise is incorporated in the direction of travel. Specifically, let there be $N$ 
particles in a volume of $V$, their equations of motion are:
\beqn
\frac{\dd  \bbr_i}{\dd t} &=& \frac{2u}{\pi} \bv(\theta_i)
\\
\frac{\dd  \theta_i}{\dd t} &=& -\frac{\pp U}{\pp \theta_i}(\bR, \Theta)+\sqrt{2D}\eta_i(t)
\eeqn
where $1 \leq i \leq N$, $\bR \equiv (\bbr_1, \ldots, \bbr_N)$, $\Theta \equiv (\theta_1, \ldots, \theta_N)$, $\bv(\theta)\equiv (\cos \theta, \sin \theta)$, and the 
noise is assumed to be Gaussian characterized by the following moments:
\beq
\la \eta_i(t) \ra =0 \sep \la \eta_i(t)\eta_j(t') \ra = \delta_{ij} \delta(t-t')
\eeq
Moreover, the alignment interaction is assumed to be of very short range and thus can be approximated by a delta function:
\beq
U(\bR, \Theta) =- \frac{g}{\pi}\sum_{i<j}\delta^{(2)}(\bbr_i -\bbr_j)  \cos(\theta_i-\theta_j)\ .
\eeq
If we denote the probability distribution of the density of particles in the state $(\bR, \Theta)$ at time $t$ by $f(t, \bR, \Theta)$, then the Fokker-Planck equation 
corresponding  to the system is \cite{Zwanzig_B01}:
\beqn
\nonumber
\frac{\pp f}{\pp t} &=&  \sum_i \bigg\{ D\frac{\pp^2 }{\pp \theta_i^2}f  -\frac{2u}{\pi}\nabla_{\bbr_i} \cdot [\bv(\theta_i) f] \bigg\}
\\
\label{feq}
&&+\frac{g}{\pi} \sum_{i< j}
\frac{\pp}{\pp \theta_{i}} \left[ \delta^{(2)}(\bbr_i -\bbr_j) 
  \sin(\theta_i-\theta_j)  f \right] \ .
\eeqn

Naturally, we are not interested in all the information captured by $f$, and we will from now on focus on the single-particle density function, $\rho$,
\beq
\nonumber
\rho(\bbr_1, \theta_1) = \frac{(N!)\int \dd r_2 \cdots \dd r_N \dd \theta_2 \cdots \dd \theta_N f(\bR, \Theta)}{(N-1)!}
\ .
\eeq
From \eq (\ref{feq}), we can express 
$\rho$ in terms of the two-particle density function $\rho^{(2)}$:
\begin{small}
\beqn
\nonumber
\frac{\pp \rho(\bbr,\theta)}{\pp t} =
D \frac{\pp^2 \rho(\bbr,\theta)}{\pp \theta^2} 
-\frac{2u}{\pi} \left[
\cos \theta \frac{ \pp \rho(\bbr,\theta)}{\pp x} +\sin \theta \frac{ \pp \rho(\bbr,\theta)}{\pp y} \right]
\\
+\frac{g}{\pi}\frac{\pp }{\pp \theta} \bigg[
 \int \dd \theta' \int \dd r' \delta^{(2)}(\bbr -\bbr') 
\sin (\theta-\theta')
\rho^{(2)}(\bbr,\theta,\bbr',\theta')
\bigg] \ .
\eeqn
\end{small}
where
\beq
\nonumber
\rho^{(2)}(\bbr_1, \theta_1,\bbr_2,\theta_2) =  \frac{(N!)\int \dd r_3 \cdots \dd r_N \dd \theta_3 \cdots \dd \theta_N f(\bR, \Theta)}{(N-2)!}
\ .
\eeq
The above manipulation is akin to the BBGKY hierarchy formalism \cite{Huang_B88}.
To continue with our analytical treatment, we will ignore the second ordered correlation and adopt the {\it product distribution} assumption: $\rho^{(2)}(\bbr, 
\theta,\bbr',\theta') = 
\rho(\bbr, \theta)\rho(\bbr', \theta')$. This assumption is similar to the molecular chaos assumption in the context of Boltzmann equation, and is also adopted in 
\cite{Bertin_PRE06, Bertin_JPA09}.

By Fourier transforming the above equation with respect to the angular variable, $\theta$, we have:
\beqn
\nonumber
\pp_t\hrho_n(\bbr) &=&
-Dn^2 \hrho_n(\bbr)
-u  \Big[ \pp_x \big(\hrho_{n+1}(\bbr)+\hrho_{n-1}(\bbr)\big) 
\\
\nonumber
&&
+\ii \pp_y \big( \hrho_{n-1}(\bbr) - \hrho_{n+1}(\bbr)\big)\Big]
\\
\label{main_eq}
&& -g n\Big[\hrho_{-1}(\bbr) \hrho_{n+1}(\bbr) 
-
\hrho_{1}(\bbr) \hrho_{n-1}(\bbr) \Big] 
\eeqn
where $\rho(\bbr, \theta) = \sum_{n=-\infty}^{\infty} \hrho_n(\bbr)\ee^{-\ii n \theta}$.

Since $\hrho_n(\bbr)$ are complex, we will denote them by $a_n(\bbr) +\ii b_n(\bbr)$ where $a_n$ and $b_n$ are real functions. In relation to the 
original density function, we have
\beqn
\label{Fourier}
\rho(\bbr, \theta) &= & \sum_{n\in {\bf Z}} [a_n(\bbr) +\ii b_n(\bbr) ]\ee^{-\ii n \theta}
\\
\nonumber
&=&
a_0(\bbr) + 2\sum_{n>1} \Big[a_n(\bbr) \cos(n \theta) +b_n(\bbr) \sin(n \theta) \Big] 
\eeqn
where for the second equality, the following conditions for the $a_n$ and $b_n$ have been employed:
\beq
a_n = a_{-n}
\sep
b_n = -b_{-n} \ ,
\eeq
which are due the fact that $\rho$ is real.
Writing \eq \ref{main_eq} in terms of the $a_n$ and $b_n$, we have for $n \in {\bf Z}$,
\begin{widetext}
\beqn
\label{LSA1}
\pp_t a_n &=&
-Dn^2 a_n -u\Big[ \pp_x (a_{n+1}+a_{n-1}) -\pp_y(b_{n-1}-b_{n+1}) \Big]
-g n\Big[a_{1}(a_{n+1}-a_{n-1})
 +b_1 (b_{n-1}+b_{n+1})\Big]
\\
\label{LSA2}
\pp_t b_n
&=&
-Dn^2 b_n 
-u\Big[ \pp_x (b_{n+1}+b_{n-1}) +\pp_y(a_{n+1}-a_{n-1}) \Big]
-g n\Big[a_{1}(b_{n+1}-b_{n-1})
-b_1 (a_{n+1}+a_{n-1})\Big] \ .
\eeqn
\end{widetext}
Note that the arguments $(t, \bbr)$ in $a_n$ and $b_n$ are omitted in the above equations to ease notation.

\section{Mean-field approximation}
\label{sMF}
To avoid having to deal with the above infinite set of differential equations, we will sort to truncate the number of differential equations to be considered. To do so, 
we first study the system in a mean-field manner \cite{MFref}, i.e., we set $\hrho_n(t,\bbr) =\hrho_n(t)$ for all $\bbr$. \eqs (\ref{LSA1}) and (\ref{LSA2}) then become
\beqn
\nonumber
\frac{\dd a_n}{\dd t} &=&
-Dn^2 a_n -g n\Big[a_{1}(a_{n+1}-a_{n-1})
\\
\label{MF1}
&&
 +b_1 (b_{n-1}+b_{n+1})\Big]
\\
\nonumber
\frac{\dd b_n}{\dd t} 
&=&
-Dn^2 b_n -g  n\Big[a_{1}(b_{n+1} -b_{n-1})
\\
\label{MF2}
&&
 -b_1(a_{n+1}+ a_{n-1})\Big] \ .
\eeqn
Note that $\dd a_0 /\dd t=0$ due to the fact that $a_0$ corresponds to the 
overall density of the system, which does not change.

Let us assume that the $b$ modes are not excited at $t=0$ and so we need only focus on the $a$ modes. By inspecting 
\eq \ref{Fourier}, we see that the omission of the $b$ modes is the same as focusing only on angular perturbation of the form $\cos (\theta)$, i.e., the particles are more likely to be heading in the positive $x$ direction. With this simplification, the first three modes are of the form:
\beqn
\frac{\dd a_1}{\dd t}  &=& g a_0a_{1} - (D a_1 + g a_1a_2)
\\
\frac{\dd a_2}{\dd t} &=& 2g a_{1}^2 - (4D a_2 + g a_1a_3)
\\
\frac{\dd a_3}{\dd t} &=& 3g a_1a_{2} - (9D a_3 + g a_1a_4) \ .
\eeqn
At the onset of collective motion (CM) from a spatially and angularly homogeneous system, we expect that $|a_n| \ll 1$ for $n>1$. Let us define $\epsilon$ as 
$\max_{n>1} |a_n|$ at the onset of CM, we see that only $\dd a_1 /\dd t$  is of order $\epsilon$ while all the time-derivatives for the higher order modes are 
of order $\epsilon^2$.
Furthermore, the coefficients associated with the damping term $D$ for the $n-$th modes scale with $n^2$, which further suggests that only the lower order modes are important. 
Another corroborating evidence is from \cite{Bertin_PRE06, Bertin_JPA09} where the authors employed their {\it scaling ansatz}, which is supported by their numerical simulations, to argue that the higher order modes are indeed negligible at the onset of CM. 
Based on all these reasons, we will truncate the original dynamical equations, \eq (\ref{main_eq}), by omitting all $\hrho_n$ for $n>2$. Focusing again only on the $a$ modes, we have
 \beqn
\frac{\dd a_0}{\dd t} &=&0
\\
\frac{\dd a_1}{\dd t} &=& (-D+ga_0-ga_2)a_1
\\
\frac{\dd a_2}{\dd t} 
&=&
-4Da_2+2ga_1^2  \ .
\eeqn
A simple fixed point analysis on the above equations indicate that the 
existence of non-zero fixed point for $a_1$ and $a_2$ is only  possible when
\beq
\label{MFcond}
g  a_0 -D >0 \ .
\eeq
This condition has previously been derived in \cite{Peruani_EPJST08}. Expectedly, the above condition indicates that collective motion is facilitated by having 
strong interaction ($g$), high particle density ($a_0$) and weak noise ($D$).

\section{Linear stability analysis}
\label{sLSA}
We now continue with our truncation approximation, but with the spatial variable re-installed into \eqs (\ref{MF1}) and (\ref{MF2}). Before we start to analyze the set of differential equations,  we note that by inspecting \eqs (\ref{LSA1}) and (\ref{LSA2}), we see that the $a_{n\geq 1}$ and $b_{n \geq 1}$ modes are coupled exclusively to 
 different spatial dimensions -- the $x$ and $y$ dimensions respectively. In other words, 
if the system is 
initially homogeneous in the $x$ dimension, then the $x$ dimension will remain homogeneous, and vice versa. We will therefore, as in the previous section, assume that the $b$ modes are not excited and focus only on the
$a$ modes. With this simplification, we arrive at the following dynamical equations:
\beqn
\label{Lalpha}
\pp_t \alpha &=& -2u\pp_x \beta
\\
\label{Lbeta}
\pp_t \beta
&=&
-D \beta
-u\pp_x(\alpha +\gamma) +g \beta(\alpha-\gamma)
\\
\label{Lgamma}
\pp_t \gamma
&=&
-4D \gamma
-u\pp_x\beta  +2g \beta^2
 \ ,
\eeqn
where we have used the Greek letters $\alpha$, $\beta$ and $\gamma$ to denote $a_0$, $a_1$ and $a_2$ respectively. 

The fixed point in the homogeneous phase corresponds to $\alpha = 1$, $\beta =0$ and $\gamma=0$ where we have set the unit length in such a way that the density of the particles is one. We now perform linear stability 
analysis on this fixed-point by considering the linear response of the system to a small perturbation  of the form:
\beqn
\alpha &=& 1 + A \ee^{\lambda t+\ii q y}
\\
\beta &=& B \ee^{\lambda t+\ii q y}
\\
\gamma &=& C\ee^{\lambda t+\ii q y}
\eeqn
where $A,B,C \ll 1$ and $q$ is an arbitrary frequency. Substituting the above into \eqs (\ref{Lalpha}) and (\ref{Lbeta}) gives the following condition on $\lambda$:
\beq
\frac{\lambda (\lambda+D-g)(\lambda+4D)}{3\lambda +8D} = -u^2q^2\ ,
\eeq
which indicates that $\lambda >0$ if and only if $g >D$. In other words, we have recovered the condition found in our previous mean-field analysis (\cf \eq 
(\ref{MFcond})).

Although this result is consistent with what we found in the previous section, pieces of the picture at the onset of CM are still lacking. For instance, the phase 
transition condition found here does not depend on the speed of the particle $u$. This is unsatisfactory because we know that long range order would not be possible 
if $u=0$ \cite{Toner_PRL95, Toner_PRE98}. Moreover, it is desirable to see how the coupling between the spatial and temporal dimension affects the rise of the excited 
mode $\beta$. To gain insight in these questions, we will go beyond the linear stability regime and analyse the dynamical equations with perturbative method in the next 
section.

\begin{figure}\caption{
The density profiles of $\alpha$, $\beta$ and $\gamma$ at different times obtained by numerically solving the set of differential equations in \eqs (\ref{alpha}) to (\ref{gamma}), with the following parameters: $u = 1/\sqrt{3}$, and $D=g = \xi=\sigma =0.1$. 
}
\label{density}
\begin{center}
\includegraphics[scale=.45]{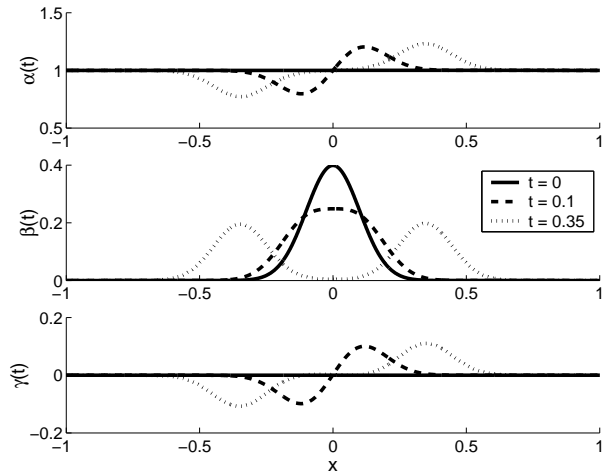}
\end{center}
\end{figure}

\section{Beyond linear stability analysis}
\label{sLast}
In this section, we will study \eqs (\ref{Lalpha}) to (\ref{Lgamma}) perturbatively. Specifically, we will assume that $D, g \ll 1$, in the units of distance and 
time set by having $\alpha (t=0,x) =1$ and $u=1/\sqrt{3}$. Physically, these assumptions amount to limiting our discussion to the regime where the particles' angular fluctuations and interaction strength are small.
Furthermore, since we are primarily interested in the dynamics at the onset of CM, we will assume that 
$g/ D$ is of order unity (\cf \eq (\ref{MFcond})). 
These assumptions allow us to employ $D$ (or equivalently, $g$) as the expansion parameter in our perturbative treatment.
In contrast to the linear stability analysis in the previous section
where the perturbation magnitude is assumed to be small enough that the nonlinear term is negligible, the perturbative approach adopted here allows us to study the effects of the
nonlinear term on the dynamics.

In the aforementioned units, \eqs (\ref{Lalpha}) to (\ref{Lgamma}) are:
\beqn
\label{alpha}
\pp_t \alpha &=&
-\frac{2}{\sqrt{3}}  \pp_x \beta
\\
\label{beta}
\pp_t \beta &=&
-D \beta 
-\frac{1}{\sqrt{3}}   \pp_x (\alpha+\gamma) +g \beta (\alpha -\gamma)
\\
\label{gamma}
\pp_t \gamma &=&
-4D \gamma
-\frac{1}{\sqrt{3}}   \pp_x \beta +2g \beta^2
\ .
\eeqn
We now expand $\alpha$ as:
\beq
\alpha = \alpha_0 + D \alpha_1 + \cO(D^2) \ ,
\eeq
and similarly for $\beta$ and $\gamma$.
The zero-th order (in $D$) terms follow the following differential equations:
\beqn
\pp_t \alpha_0 &=&
-\frac{2}{\sqrt{3}}  \pp_x \beta_0
\\
\pp_t \beta_0 &=&
-\frac{1}{\sqrt{3}}   \pp_x (\alpha_0+\gamma_0)
\\
\pp_t \gamma_0 &=&
-\frac{1}{\sqrt{3}}   \pp_x  \beta_0
\ .
\eeqn
The above set of differential equations can be solved by employing the Laplace-Fourier Transform method. For the initial conditions of $\alpha_0(t=0,x) =1$,
$\beta_0(t=0,x)= \xi \ee^{-y^2/(2\sigma^2)}/\sqrt{2\pi }\sigma$ and $\gamma_0(t=0,x) = 0$. The solutions are:
\beqn
\label{Z1}
\alpha_0 &=& 1+\frac{\xi}{\sqrt{6\pi}\sigma}\bigg[ U^--U^+\bigg]
\\
\label{Z2}
\beta_0
&=&\frac{\xi}{2 \sqrt{2\pi}\sigma}\bigg[ U^-+U^+ \bigg] 
\\
\label{Z3}
\gamma_0 &=& \frac{\xi}{2 \sqrt{6\pi}\sigma}\bigg[ U^--U^+\bigg]
\ ,
\eeqn
where 
\beq
\cA^\pm = \exp \left(-\frac{(x\pm t)^2}{2\sigma^2} \right) \ .
\eeq
In other words, a Gaussian perturbation in $\beta$ at $t=0$ splits into two Gaussian distributions traveling in opposite directions with unit speed (as a result of 
setting $u$ to $1/\sqrt{3}$). The perturbation also induces in $\alpha$ and $\gamma$ two solitary  waves in the form a Gaussian distribution traveling with unit speed in the positive direction, and 
an inverted Gaussian density wave traveling in the opposite direction (\cf \fig \ref{density}). This is akin to the stripe traveling wave pattern found in the CM phase
\cite{Gregoire_PRL04,Bertin_JPA09}.

The differential equations governing the first-order terms are:
\beqn
\label{A1}
\pp_t \alpha_1 &=&
-\frac{2}{\sqrt{3}}  \pp_x \beta_1
\\
\label{B1}
\pp_t \beta_1 &=&
-\beta_0 
-\frac{1}{\sqrt{3}}   \pp_x (\alpha_1+\gamma_1) +\frac{g}{D} \beta_0 (\alpha_0 -\gamma_0)
\\
\label{C1}
\pp_t \gamma_1 &=&
-4 \gamma_0
-\frac{1}{\sqrt{3}}   \pp_x \beta_1 +\frac{2g}{D} \beta^2_0
\ .
\eeqn
where $\alpha_0$, $\beta_0$ and $\gamma_0$ above are now given by \eqs (\ref{Z1}) to (\ref{Z3}). 
The initial conditions for the above equations are: $\alpha_1(t=0,x)=\beta_1(t=0,x)=\gamma_1(t=0,x)=0$. 

\begin{figure}\caption{
The temporal evolutions of (a) $A$ (\cf \eq (\ref{A})) and (b) $B$ (\cf \eq (\ref{A})), obtained by numerically solving the set of differential equations in \eqs 
(\ref{alpha}) and (\ref{gamma}), with $D=g = \xi=0.1$. (c) The zoom-in plot of $B(t)$ at small time with the three curves corresponding to the theoretical expressions 
given in \eq (\ref{T1}).
}
\label{AB}
\begin{center}
\includegraphics[scale=.45]{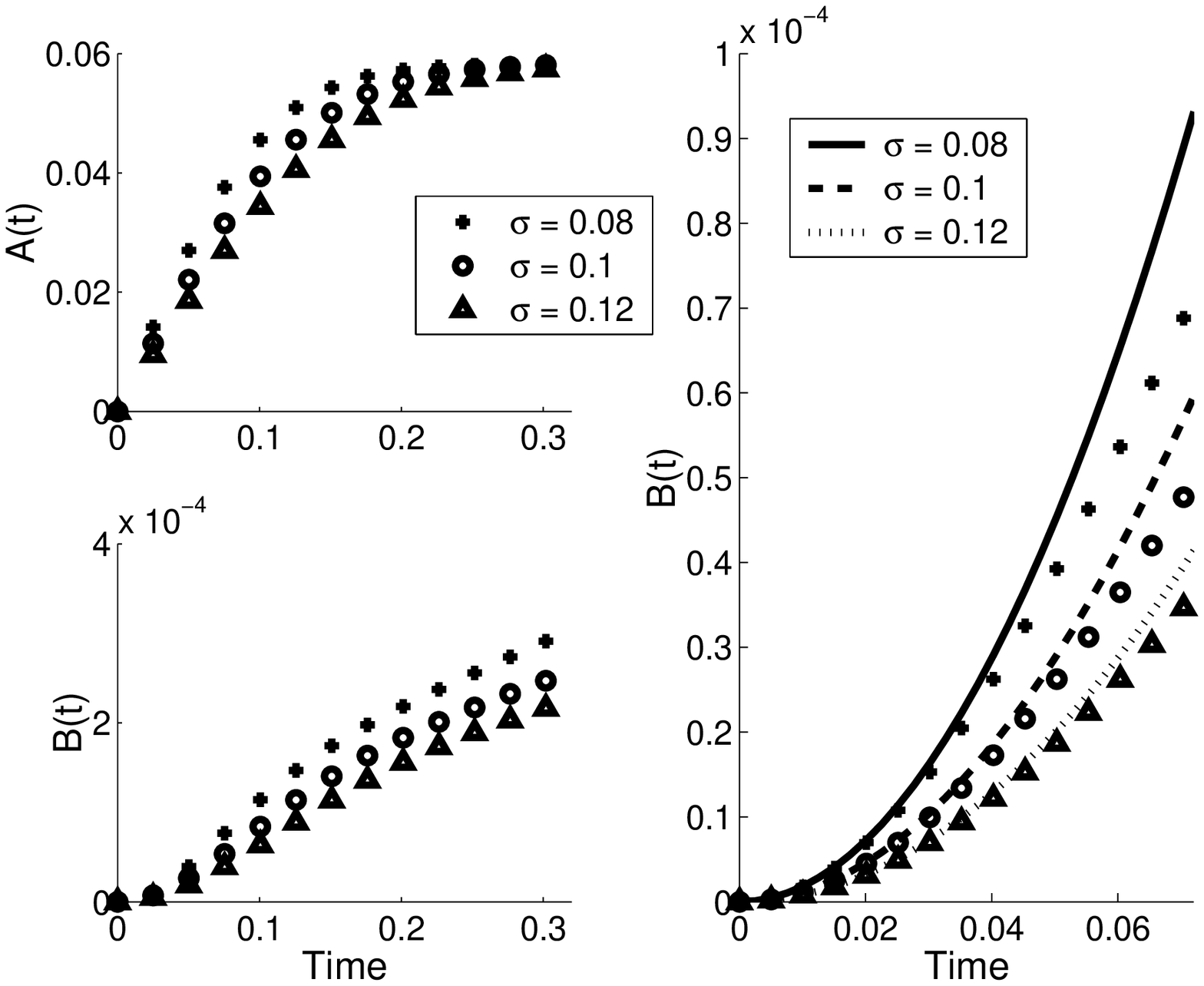}
\end{center}
\end{figure}
\begin{figure}\caption{
The temporal evolutions of (a) $A$ (\cf \eq (\ref{A})) and (b) $B$ (\cf \eq (\ref{A})), obtained by numerically solving the set of differential equations in \eqs 
(\ref{alpha}) and (\ref{gamma}), with $ g = b=\sigma=0.1$. The zoom-in plot of $B(t)$ at small time with the three curves corresponding to the theoretical 
expressions given in \eq (\ref{T1}).
}
\label{AB2}
\begin{center}
\includegraphics[scale=.45]{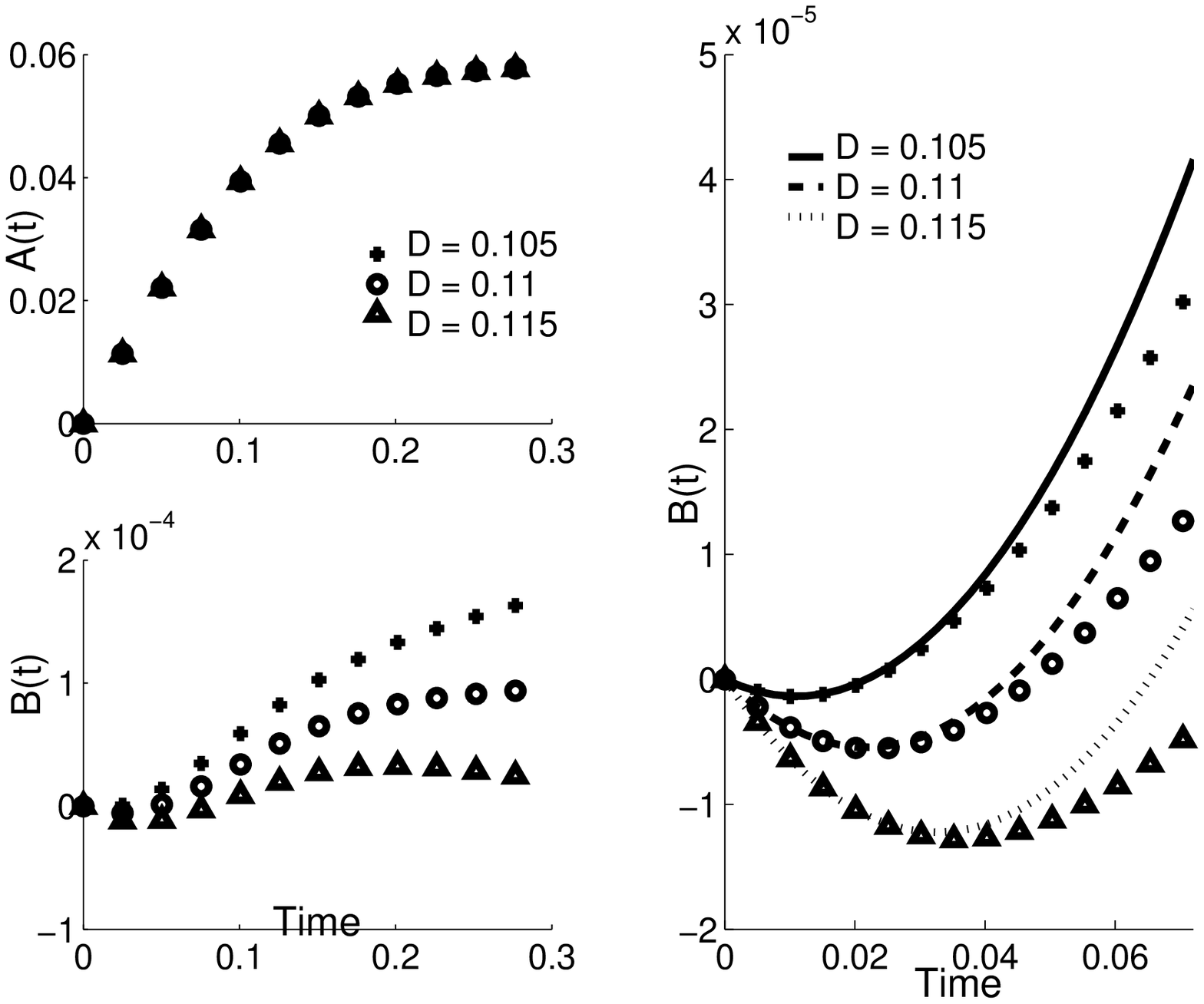}
\end{center}
\end{figure}

Before we attempt to study the above set of differential equations, let us look for some meaningful quantities that quantifies the effect of the initial perturbation. For instance, the temporal 
evolution of total increase in $\beta_1$ due to the initial perturbation can be obtained from \eq (\ref{B1}):
\beqn
\nonumber
\pp_t \left(\int_{-\infty}^\infty \dd y \beta_1(t,y) \right) &=&
\int_{-\infty}^\infty  \dd y\left( \frac{g}{D}\alpha-1\right) \beta
\\
&=& \xi\left( \frac{g}{D}-1\right)
\eeqn
This indicates that when summed over the whole space, the $\beta$ mode is amplified  only if $g >D$. This is again consistent with the result we obtained in 
Sect.~\ref{sMF} and Sect.~ \ref{sLSA}.

Besides the above quantity, the following two quantities are also of interest:
\beq
\label{A}
A(t) \equiv  \int_{0}^\infty \dd x \alpha(t,x) \ ; \
B(t) \equiv \int_{0}^\infty \dd x \big[ \beta(t,x) -\beta(0,x)\big]\ .
\eeq
Namely, $A$ and $B$ correspond to the responses in the density ($\alpha$) and in the $\beta$ mode {\it in the direction} of the angular perturbation. These are in fact 
arguably better quantities to consider as they capture the directional nature of the perturbation.
From \eq (\ref{B1}), we have:
\beqn
\frac{\dd B(t)}{\dd t} &=& 
 \frac{\xi( g-D)}{2}+ \frac{g \xi^2}{8 \sqrt{3\pi} \sigma} \erf\left(\frac{t}{\sigma} \right) 
\\
&&+
\frac{D}{\sqrt{3}} [ \alpha_1(t,x=0)+\gamma_1(t,x=0)]
\ .
\eeqn
In the appendix, we demonstrate that
\beqn
\alpha_1(t,y=0)+\gamma_1(t,y=0) = \frac{g \xi^2}{\pi \sigma^2} t +\cO(t^3) \ .
\eeqn
In other words, up to order $\cO (t^3)$, we have
\beq
\label{T1}
B(t) = \frac{b( g-D)t}{2}+  \frac{5 g t^2}{8\sqrt{3} \pi } \left(\frac{\xi}{\sigma}\right)^2\ .
\eeq
The third term in the right hand side above highlights the importance of the term $\xi/\sigma$, especially when 
$D\simeq g$. \fig \ref{AB}(a) and (b) display the temporal evolutions of $A(t)$ and $B(t)$ by solving \eqs (\ref{alpha}) and (\ref{beta}) numerically in the case of 
$D= g$. They clearly show the amplification of the initial perturbation, which we have taken as an indication for the onset of collective motion in longer time. 
\fig \ref{AB}(c) demonstrates that at short time, the dynamics is well described by the expression in \eq (\ref{T1}). Furthermore, due to the positive second term in the 
R.H.S. above, the formula for $B(t)$ suggests that there is a possibility of perturbation amplification even if $D>g$, e.g., when $\xi/\sigma \gg 1$. This is indeed 
shown to be the case in \fig \ref{AB2}(a) and (b), where amplification of the perturbation is seen for $D/g \sim 1.1$. In other words, a sharp perturbation in the 
angular domain is able to induce collective motion even if the density is below the phase transition threshold as obtained in the mean-field model. 

If we now restore the speed, $u$, and the initial density, $c$, where $c= \alpha(t=0,x)$, into \eq (\ref{T1}), we have
\beq
B(t) = \frac{bc( cg-D)t}{2}+ \frac{ 5g uct^2}{8\pi } \left(\frac{\xi}{\sigma}\right)^2\ .
\eeq
Note that the speed of the particles only appears in the second term above, which is positive. Hence, the above formula suggests that the particle speed has a net effect 
of amplifying the initial perturbation and thus facilitating collective motion transition.

\begin{figure}\caption{
The evolution of $B(t)$ obtained by numerical simulating the differential equations \eqs (\ref{LSA1}) and (\ref{LSA2}) with the initial condition discussed in \sect \ref{sLast}. The parameters are $g = D= \xi =\sigma =0.1$. In the simulations, only the 0-th to the $m$-th $a$ modes are included. In other words, we assume that $a_{n>m} =0$ and $b_{n\in {\bf Z}}=0$.
} 
\label{acc}
\begin{center}
\includegraphics[scale=.35]{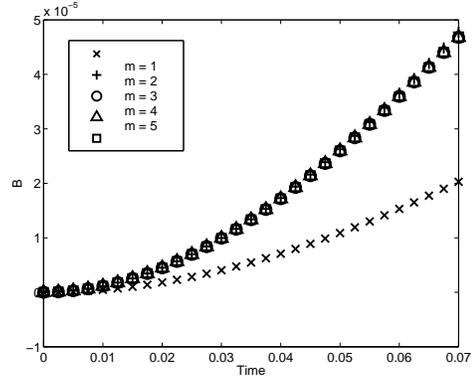}
\end{center}
\end{figure}

As a verification on the validity of our approximation adopted in our analytical calculations, we numerically simulate \eqs (\ref{LSA1}) and (\ref{LSA2}) with higher order modes included and find that there  are no  discernible differences for the parameter range investigated in this work (\cf \fig \ref{acc}).

\section{Conclusion}

In summary, starting with a Fokker-Planck equation for a minimal model of CM, we derived a set of three coupled differential equations that describes the system at the 
onset of CM. We then studied the equations with mean-field, linear stability, perturbative and numerical analyses, and found that if 
an angular perturbation is large in magnitude and highly localized in space, it will be amplified and thus serves as an indication of the onset of collective motion. Our calculations also demonstrate the importance of particle speed for collection motion transition. As a result, it is indicative that the critical point for CM may depend 
on the speed $u$, the perturbation magnitude $b$ and the perturbation wavelength $\sigma$. This is in contrast to the mean-field and linear stability analyses 
where only the hydrodynamic, or infinite-wavelength, mode dictates the onset of CM. Our results therefore highlights the importance of incorporating the nonlinear term into the analysis.

The main limitation of this work is on the approximation adopted -- the omissions of higher order modes. While we believe that such an approximation is appropriate at 
the onset of CM, it would be highly desirable to have a systematic method to incorporate the higher order modes into the dynamics. Besides the consideration of the 
higher order modes, singular perturbation method would also be needed to investigate the long-time behaviour of the system \cite{footnote}. We believe that these 
aspects would constitute two promising directions for future investigation.

\appendix*
\section{}
We are unable to solve the set of differential equations shown in \eqs (\ref{A1}) to (\ref{C1}) analytically. But since only the leading orders in $x$ and $t$ are of interests, we will replace the $U^\pm$ in $\alpha_0, \beta_0, \gamma_0$ (\cf \eqs (\ref{Z1}) to (\ref{Z3})) by 
\beq
\tilde{U}^\pm \equiv 1 - \frac{(x\pm t)^2}{2 \sigma^2} + \left[\frac{(x\pm t)^2}{2 \sigma^2}   \right]^2 \ .
\eeq
With this simplification, the differential equations can be solved by the Laplace-Fourier Transform method and the relevant results are:
\beq
\alpha_1(t,y=0) = \cO(t^3) \sep
\gamma_1(t,y=0) = \frac{g \xi^2}{\pi \sigma^2} t + \cO(t^3)\ .
\eeq

\end{document}